\documentclass[twocolumn,preprintnumbers,amsmath,amssymb]{revtex4}
\usepackage{graphicx}
\usepackage{dcolumn}
\usepackage{bm}

\begin{document}

\bibliographystyle{alpha}
\title{ The Influence of Cylindrical Inclusions on the Stability of a
Directionally Solidified Interface}
\author{ Layachi  Hadji}
\email{Lhadji@bama.ua.edu}
\affiliation{The University of Alabama,
Department of Mathematics,
Tuscaloosa, Alabama 35487}
 
\date{\today}
\maketitle

It is well known that the presence of an inclusion in the melt near
a solidifying front induces a local deformation in the latter provided that
the melt's thermal conductance differs from that of the inclusion
\cite{Zub73,Aac77,Sen97,Had01}.  This local interfacial deflection  is
caused by the modification of the thermal gradient in the melt near the
particle. The long time evolution of this local disturbance of the interface
has been investigated for the case of a spherical particle in a pure
substance \cite{Had03}, and in a binary alloy \cite{Had04}.
It is discovered that, provided that the particle-interface distance falls
below a critical value, the induced perturbation grows and destabilizes the
whole solid-liquid interface. This newly uncovered morphological instability 
is manifested only for some combination of the physical and processing 
parameters, and its onset is attributed to the reversal of the thermal gradient
in the liquid gap between the particle and the interface \cite{Had04}.
This instability,
whose characteristic size is of the order of the particle's diameter,
occurs at pulling speeds that are below the threshold for the onset of the 
Mullins-Sekerka instability.\\

In reality, however, the inclusions have approximately cylindrical shapes.
For example, in their study involving a particulate metal matrix composite
of magnesium base alloy that is reinforced with SiC particles,
Essa {\it et. al} \cite{Ess02} state that the inclusions resemble 
cylinders whose longitudinal sections can be approximated by
rectangles of sides $a$ and $b$ ($a \ne b$). The pupose of this communication
is to extend the stability analysis that was carried out for a spherical
inclusion in \cite{Had03} to the more realistic case of a cylindrical inclusion.
We let the inclusion have the more general shape of an elliptic cylinder, i.e.
the longitudinal section is an ellipse of semi-axes $a$ and $b$ as illustrated
in figure 1. The effect of the aspect ratio $c=b/a$, which emerges in the
analysis as an important factor, will be examined.\\

\begin{figure}
\includegraphics[height=2.4in]{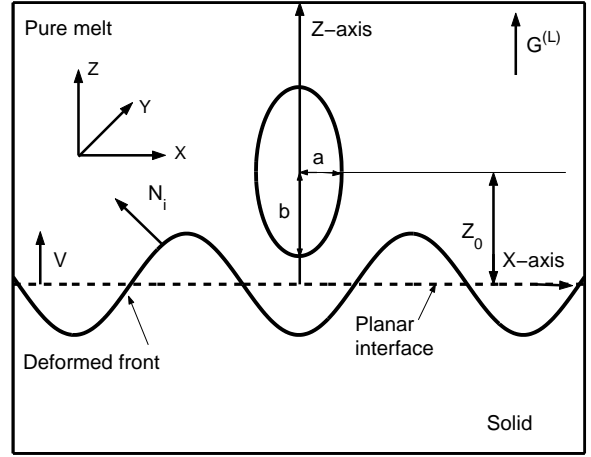}
\caption{\label{fig:epsart}A schematic diagram of a vertical solidification
setup of a pure melt in which is embedded a cylindrical inclusion of aspect
ratio $c=b/a$. The distance from the inclusion's center to the planar
interface is $Z_0$. The cylinder's axis of symmetry is parallel to the
$Y-$axis, and the planar interface coincides with the $X-axis$.
The front's growth rate is $V$, $G^{(L)}$ is the imposed thermal gradient, 
and ${\bf N}_i$ is the normal to the interface pointing into the melt.}
\end{figure}

Consider a system in which an inclusion is immersed in a bath of pure melt
that is undergoing directional solidification. We ignore the gravitational
effects and assume that the inclusion is both solid, i.e. nondeformable, and
insoluble in the melt. Let the inclusion's position be such that the
distance from its center to the planar solid-liquid interface is $Z_0$, and
its axis of symmetry is parallel to the $Y-$axis. The problem, being 
invariant with respect to the $Y-$direction, is thus two-dimensional.
The physical process is governed soleley by heat diffusion in the melt,
solid, and inclusion with appropriate conditions at the interface, at the
inclusion's surface, and at infinity. Let $T^{(q)}(X,Z)$ represent the
temperature in phase $(q)$, where the superscript $(q)$ denotes symbolically
the liquid phase, $(q=L)$, the solid phase, $(q=S)$, and inclusion, $(q=I)$,
and the couple $(X,Z)$ represents the Cartesian coordinates of a point in
the $XZ$-plane of a right-handed coordinate system
 relative to which the solid-liquid interface
is at rest at $Z=0$, i.e. the $X$-axis. The melt occupies the top region,
$Z>0$ as shown schematically in Fig. 1.  The equation of the cylindrical 
inclusion, whose center is at the point $(0,Z_0)$, is given by 
\begin{equation}
\frac {X^2}{a^2} + \frac {(Z-Z_0)^2}{b^2} =1,
\end{equation}
where $a$ and $b$ are the semiaxes in the $X$ and $Y$ directions, respectively..
The nondimensionalization that is adopted here makes use of
$a$, $a/V$, and the melting point $T_m$ as scales for
length, time, and temperature. The following dimensionless system
of equations and corresponding boundary conditions emerges,
\begin{equation}
\epsilon \lambda^{(q)} ( \frac {\partial T^{(q)}}{\partial t} -
\frac {\partial T^{(q)}}{\partial z} ) =
{\Delta} T^{(q)},
\label{eq:one}
\end{equation}

\begin{equation}
T^{(S)} = T^{(L)} = 1 - \sigma \kappa, \quad {\mbox{at}}
 \quad z=0,
\label{eq:three}
\end{equation}

\begin{equation}
{\cal S} v_{n} = k {\partial T^{(S)} \over \partial n_i}
- {\partial T^{(L)} \over \partial n_i}, \quad {\mbox{at}} \quad z=0,
\label{eq:four}
\end{equation}

\begin{equation}
T^{(L)} = T^{(I)}, \quad
\alpha {\partial T^{(P)} \over \partial n_P} = {\partial T^{(L)} \over \partial
n_P}, \quad {\mbox{on}} \quad \partial I 
\label{eq:six}
\end{equation}

\begin{equation}
{\partial T^{(S)} \over \partial z} \rightarrow G_S, \quad 
{\partial T^{(L)} \over \partial z} \rightarrow G_L, \quad {\mbox{as}} \quad
z \rightarrow -\infty \quad {\mbox{and}} \quad \infty.
\end{equation}

The symbols that appear in Eqs.(2)-(6) are defined in the following:
$\Delta$ stands for the two-dimensional Laplacian, $\partial_{xx}
+\partial_{zz}$, $\epsilon = aV/D^{(L)}$  is the Peclet number, and
$\lambda^{(q)} = D^{(q)}/D^{(L)}$,
where $D^{(q)}$ is the thermal diffusion coefficient in phase $(q)$.
Equation (3) describes the continuity of temperature at the interface, 
with an interface equilibrium temperature that accounts for the 
Gibbs-Thomson effect. 
The parameter $\sigma = \sigma_{SL}/a L$ is the surface energy parameter, with
$\sigma_{SL}$ being the interface excess free energy, $L$ is the latent
heat of fusion per unit volume, and $\kappa$ is the front's curvature 
considered to be positive when the solid bulges into the melt. Equation (4)
describes the heat balance at the interface, where
 $S=LVa/(T_m k^{(L)})$ is the Stefan number, $v_n$ is the normal
interface velocity, $k = k^{(S)}/k^{(L)}$, and $\partial/\partial n_i$ is the
normal derivative at the solid-liquid interface. Equation (5) represents 
the continuity of temperature and of the heat flux on the inclusion's
boundary $\partial I$,
where $\alpha = k^{(P)}/k^{(L)}$ is the thermal conductance ratio, and
$\partial/\partial n_P$ is the normal derivative at the inclusion's surface.
Finally, Eq. (6) describes the far field conditions for the temperature,
where $G_S = a G^{(S)}/T_m$ and $G_L = a G^{(L)}/T_m$, with $G^{(L)}$ and
$G^{(S)}$ being the externally imposed (dimensional) thermal gradients in the
liquid and solid phase, respectively. \\

The average size of the reinforcements that are typically  used in the
manufacture of PMMCs is about $12\rm\,\mu\,m$ \cite{Ess02}, while the melt's
thermal diffusion coefficient is of the order $10^{-5}\rm\,m^2/s$ \cite{Kur92}.
Thus, if we assume that the growth rate is low enough, 
say $V \approx 10^{-8}\rm\,m/s$, then the corresponding
Peclet number $ \epsilon \approx 10^{-7}$.  Note that
the smallness of $\epsilon$ is due to the
fact that it is proportional to the product of the velocity and the particle's
radius, both of which are considered very small. Thus, the steady state form of 
Eq. (2) satisfies
\begin{equation}
\Delta T^{(q)} + \epsilon \lambda^{(q)} {\partial T^{(q)} \over \partial z}
= \Delta T^{(q)} + O\bigl(\epsilon\bigr).
\end{equation}
As in ref. \cite{Had04},  we could take advantage of the smallnes of $\epsilon$
and  consider the expansions, with $\epsilon \ll 1$, 
$T^{(q)} = {T_B}^{(q)} + \epsilon \theta^{(q)} + O\bigl(\epsilon^2\bigr)$
and, for the interface,
 $z = 0 + \epsilon \eta + O\bigl(\epsilon^2\bigr)$. At order unity,
the base state satisfies Laplace's equations with coupled boundary conditions
at the inclusion's surface and at the planar interface. This problem is
mathematically tractable, and the next order perturbations are easily 
calculated. We refer the interested reader to \cite{Had04} for details. 
However, for both simplicity in the calculations and clarity in the 
presentation, we set $\epsilon =0$  in the present work.  This simplification
will not alter the key qualitative features of the instability mechanism that
is the subject of this communication.\\

It is known that the thermal field in an infinite medium in which is 
embedded an inclusion of different material is given by \cite {Car59}, 
\begin{equation}
T_B^{(L)}(x,z)=1 + G_L z + {(1-\alpha) G_L \over 1 + {\cal C}_0 (\alpha-1)}
           (z + n H_0) {\cal C}_{\scriptstyle \lambda 
           (\scriptstyle x,z)},
\end{equation}
where 
\begin{equation}
{\cal C}_{\scriptstyle \lambda \scriptstyle( {x,z,n})} = {c \over 2}
\int_{\scriptstyle \lambda \scriptstyle( {x,z,n})}^{\infty}
{du \over (c^2 +u)^{3/2} (1+u)^{1/2}},
\end{equation}
$c=b/a$, $H_0=Z_0/a$, and $\lambda ({x,z,n})$ is the positive root of the
equation,
\begin{equation}
{x^2 \over 1 + \lambda \scriptstyle( {x,z,n})} 
+ { (z-nH_0)^2 \over c^2 + \lambda \scriptstyle ({x,z,n}) } =1.
\end{equation}
For an elliptic cylinder, Eq. (9) reduces to
\begin{equation}
{\cal C}_{\scriptstyle \lambda \scriptstyle (x,z,n)} =
{c \over 1-c^2} \bigl[ \sqrt{    { 1 + \lambda(x,z,n) \over
                                   c^2 + \lambda(x,z,n)}  } -1 \bigr],
\end{equation}
while for a circular cylinder, we have
$$ 
{\cal C}_{\scriptstyle \lambda \scriptstyle (x,z,n)} =
{1 \over 2[x^2 + (z - nH_0)^2] }, \quad {\mbox{and}}
$$
\begin{equation}
\lambda(x,z,n) = x^2 + (z-nH_0)^2 -1. 
\end{equation}

This solution to Laplace's equation, Eq. (8), does not, however,
satisfy the boundary condition at the planar interface, Eq.(3), i.e.
$ T_B^{(L)} \ne 1$ at $z=0$. As shown in \cite{Had01}, the isotherm at $z=0$
is deformed if the thermal conductance ratio $\alpha \ne 1$. This, in turn, 
implies that the interface profile, which conforms to the isotherm at
$z=0$, is also deformed. However, in order to conduct a linear stability 
analysis of the planar interface, we must seek solutions to Eq. (7), with
 $\epsilon =0$, and
corresponding boundary conditions, Eqs.(3)-(6), that admit a planar interface
profile. This is achieved by making use of the method of images. We refer the
interested
reader to ref. \cite{Gre98} and references therein for more details on this
method. We start by introducing an image inclusion that is centered at 
$(0, -H_0)$. On using the principle of superposition, the equation
$$
T_B^{(L)}(x,z)=1 + G_L z + {(1-\alpha) G_L \over 1 + {\cal C}_0 (\alpha-1)}
           [(z + H_0) {\cal C}_{\scriptstyle \lambda (\scriptstyle x,z)}+
$$
\begin{equation}
            (z - H_0) {\cal C}_{\scriptstyle \lambda (\scriptstyle x,z)}]
\end{equation}
not only satisfies Laplace's equation, but also the boundary condition,
$T_B^{(L)} = 1$ at $z=0$. This
situation is depicted schematically in figure 2 for the case
 of  an inclusion and its image that are both
characterized by a thermal conductance ratio $\alpha >1$.
According to \cite{Zub73,Aac77,Sen97,Had01},
the inclusion at $(0,H_0)$ induces a depression in the interface shape
while its image at $(0,-H_0)$ induces a bump. The depression and the bump,
being symmetrical about the $x-$axis, cancel out leaving the interface planar.
Note, however, that Eq. (13) fails to satisfy the boundary condition at the
inclusion's surface $\partial I$, i.e. Eq.(5). Thus,
 It is necessary to place another
fictitious particle at $(0,2H_0)$ and add its contribution to the thermal
field. The resulting equation will then satisfy Eq. (5) but fails 
to satisfy Eq. (3). So an image inclusion is placed at $(0,-2 H_0)$ and so
on and so forth. This process is continued indefinitely to yield,
\begin{figure}
\includegraphics[height=2.4in]{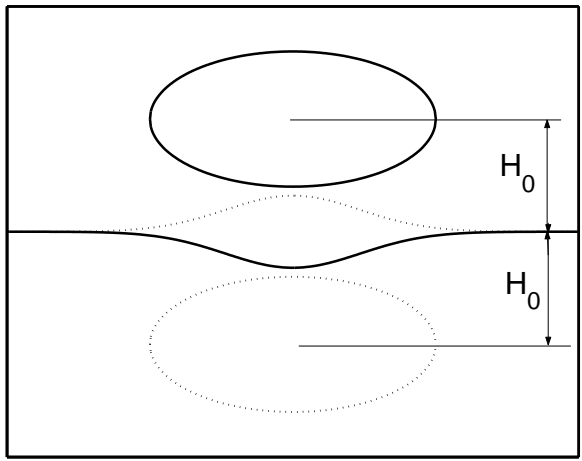}
\caption{\label{fig:epsart} A sketch of a highly heat conducting inclusion 
in the melt (continuous line) and its image (dotted line), and the
associated deformed solid-liquid interface profile.}
\end{figure}

$$ 
T_B^{(L)}(x,z)=1 + G_L z + {(1-\alpha) G_L \over 1 + {\cal C}_0 (\alpha-1)}
$$
\begin{equation}
 \times   \sum_{\scriptstyle n=-\infty\atop
          \scriptstyle n \ne 0}^{\infty} (z + n H_0) 
           {\cal C}_{\scriptstyle \lambda 
           (\scriptstyle x,z,n)}.
\end{equation}
The expression, Eq.(14), now satisfies Laplace's equation and both
boundary conditions, Eq.(3), at the planar interface, and Eq.(5) on 
$\partial I$.
The thermal fields in the inclusion and the solid phase are described by
\begin{equation}
T_B^{(P)}(z) = {G_L(z-H_0) \over 1+{\cal C}_0 (\alpha -1)},
\end{equation}
$$  
T_B^{(S)}(x,z)=1 + G_S z + {(1-\alpha) G_L \over k [1 + {\cal C}_0 (\alpha-1)]}
$$
\begin{equation}
 \times   \sum_{\scriptstyle n=-\infty\atop
          \scriptstyle n \ne 0}^{\infty} (z + n H_0) 
           {\cal C}_{\scriptstyle \lambda 
           (\scriptstyle x,z,n)}.
\end{equation}
Note that $T_B^{(S)}$ satisfies Laplace's equation, and $T_B^{(S)} =1$
at the planar interface $z=0$. The last term in Eq. (16) is necessary
in order for $T_B^{(S)}$ to satisfy the heat balance equation at the planar
interface, i.e. the planar interface moves with constant velocity; 
$T_B^{(S)}$ does not need to satisfy the boundary condition,
Eq. (5), on the particle's surface $\partial I$.\\

We now examine the linear stability of the planar state defined by
Eqs. (14)-(16) and corresponding boundary conditions at the planar interface.
We first superimpose two-dimensional, time-dependent infinitesimal
disturbances $\theta^{(L)}$, $\theta^{(S)}$ and $\eta$ upon the basic state
solutions $T_B^{(L)}$, $T_B^{(S)}$ and $z=0$, respectively in Eq. (7)
(without the convection term) and
boundary conditions, Eqs. (3)-(6). The resulting equations are then
linearized with respect to the disturbances to yield the following,
\begin{equation}
\Delta \theta^{(q)} =0,
\end{equation}
\begin{equation}
\theta^{(L)} = -G_L \eta + \sigma {\partial^2 \eta \over \partial x^2}
 - {F}(x) \eta, \quad
{\mbox{at}} \quad z=0,
\end{equation}
\begin{equation}
\theta^{(S)} = - G_S \eta + \sigma {\partial^2 \eta \over \partial x^2}
 - {{F}(x) \over k} \eta,
\quad {\mbox{at}} \quad z=0,
\end{equation}
\begin{equation}
{\partial \theta^{(L)} \over \partial z} \rightarrow 0, \quad {\mbox{as}} \quad
 z \rightarrow 
\infty, \quad \quad
{\partial \theta^{(S)} \over \partial z} \rightarrow 0, \quad {\mbox{as}} \quad
z \rightarrow -\infty,
\end{equation}
\begin{equation}
{\partial \eta \over \partial t} = k {\partial \theta^{(S)} \over
\partial z} - {\partial \theta^{(L)} \over \partial z},
\quad {\mbox{at}}  \quad z=0.
\end{equation}
The small slope approximation in the expression for the curvature, i.e.,
$\kappa \approx - \partial^2 \eta/\partial x^2$ has been used in Eqs.(18)-(19),
and
\begin{equation}
{F}(x) = {(1-\alpha) G_L \over 1 + {\cal C}_0 (\alpha -1) }
\sum_{\scriptstyle n=-\infty\atop \scriptstyle n \ne 0}^{\infty}
\Bigl[ (nH_0) {\partial {\cal C}_{\scriptstyle \lambda (x,n)} \over
               \partial z}  +  {\cal C}_{\scriptstyle \lambda (x,n)} \Bigr].
\end{equation} 
The stability problem, Eqs. (17)-(21), is solved using the Fourier transform.
We let ${\cal T} \{ f(x,z,t)\} = {\hat f}(\omega,z,t)$. 
The transformed problem is
\begin{equation}
{\partial^2 {\hat \theta}^{(q)} \over \partial z^2} - \omega^2
{\hat \theta}^{(q)} =0,
\end{equation}
with the accompanying conditions
\begin{equation}
{\hat \theta}^{(L)} = -(G_L + \sigma \omega^2) {\hat \eta} -
                      {\cal T} \{ F(x) \eta \}, \quad {\mbox{at}} \quad z=0,
\end{equation}
\begin{equation}
{\hat \theta}^{(S)} = -(G_S + \sigma \omega^2) {\hat \eta} -
                       { {\cal T}  \{ F(x) \eta \} \over k }, \quad
{\mbox{at}} \quad z=0,
\end{equation}
and with vanishing temperature gradients far away from the interface for
${\hat \theta}^{(L)}$ and ${\hat \theta}^{(S)}$. In order to make progress,
we consider the approximation ${\cal T} \{ F(x) \eta \} \approx
F(0) {\hat \eta}$ since the major contribution to $F(x)$ comes from near the
origin (see \cite{Had03} for a similar approximation for the case of a spherical
 particle), where $F(0)$ can be be approximated by the leading asymptotic
term for $c \approx 1$ (small deviation from a circular cylinder) as follows,
\begin{equation}
F(0) = \frac {(\alpha-1) G_L c}{2[1+{\cal C}_0(\alpha-1)]H_0^2}
          \sum_{\scriptstyle n=-\infty\atop
          \scriptstyle n \ne 0}^{\infty}{1 \over n^2}
+ O\bigl(c-1\bigr).
\end{equation}
The approximation is, however, exact for a circular cylinder $(c=1)$.
On using the fact
that $\sum_{n=1}^{\infty} 1/n^2=\pi^2/6$ and ${\cal C}_0 =1/(c+1)$, we obtain
\begin{equation}
F(0) \approx \frac {\pi^2 (\alpha-1) G_L c(1+c)}{12 (c+\alpha) H_0^2}.
\end{equation}

The solution of the transformed problem yields,
\begin{equation}
{\hat \theta}^{(L)} = -[ G_L + \sigma \omega^2 + F(0)] e^{- \omega z},
\end{equation}
\begin{equation}
{\hat \theta}^{(S)} = -[ G_S + \sigma \omega^2 + {F(0) \over k}] e^{\omega z}.
\end{equation}
On imposing the heat balance equation at the interface, we obtain the following
evolution equation for the interface perturbation,
\begin{equation}
{\partial {\hat \eta} \over \partial t} = \Omega(\omega) {\hat \eta},
\end{equation}
where $\Omega(\omega) = -[{\cal G} + (1+k) \sigma \omega^2 + 2 F(0)]\omega$, 
 ${\cal G} = k G_S + G_L$ is the conductivity weighted thermal gradient, and
$\Omega$ represents the growth rate of the interfacial perturbation. If
$\Omega(\omega)<0$ then any initial disturbance dies out after a long time, and 
the planar interface is stable. If $\Omega(\omega) >0$ then the opposite 
scenario holds, and the interface is unstable. The
marginal stability criterion is obtained by setting $\Omega(\omega) =0$.
A plot of the growth rate as a function of the wavenumber $\omega$ is shown
in figure 3 as function of the aspect ratio, $c$, and the inclusion-interface
distance, $H_0$. Note that the range of unstable wavenumbers, $0 < \omega <
\omega_n$, increases with the aspect ratio $c$, where $\omega_n$ is the
neutral mode, i.e. $\Omega(\omega_n)=0$. Therefore, inclusions with
larger $c$ values have a more destabilizing effect, or equivalently, the
instability is onset at a larger inclusion-interface distance.
\begin{figure}
\includegraphics[height=2.0in]{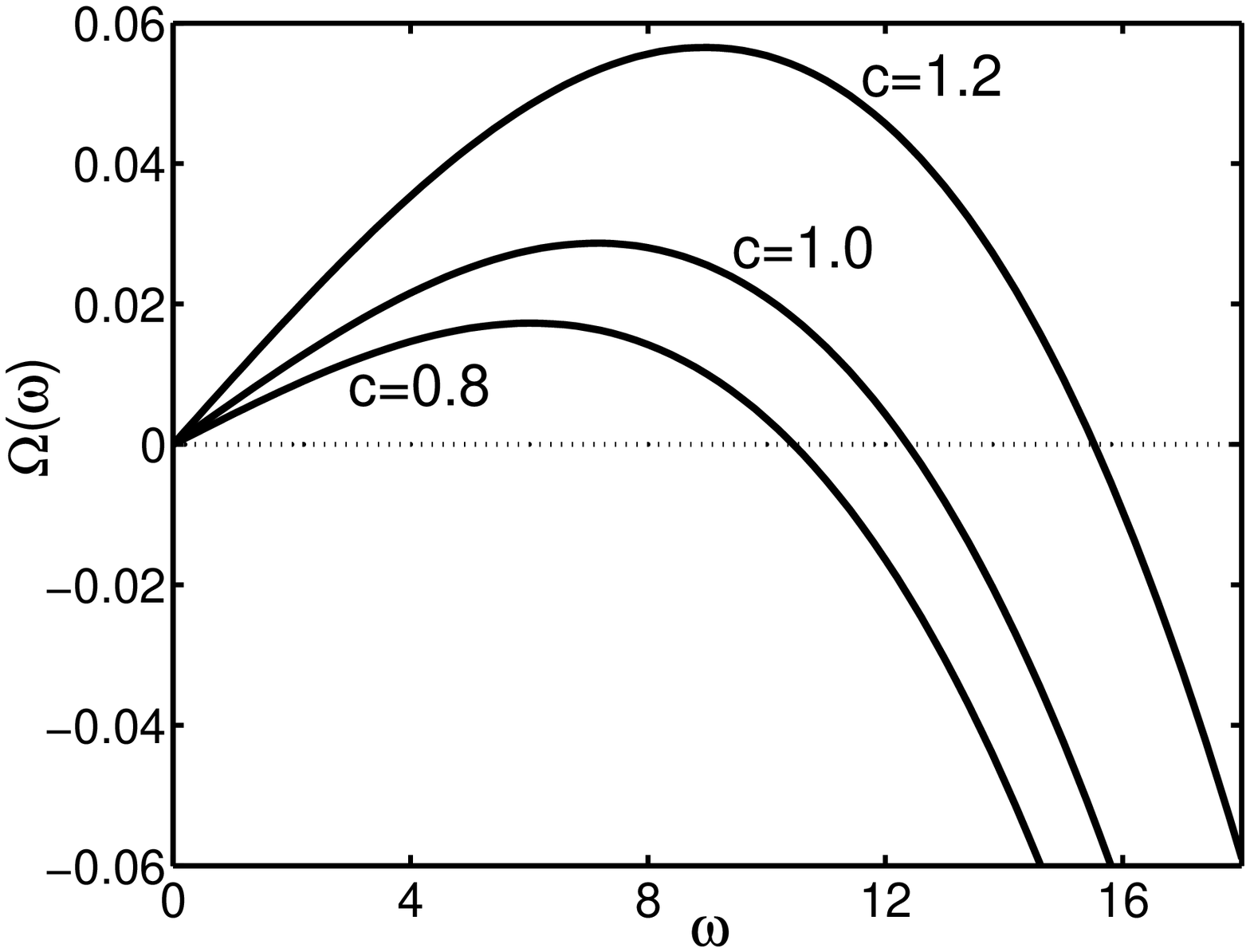}
\includegraphics[height=2.0in]{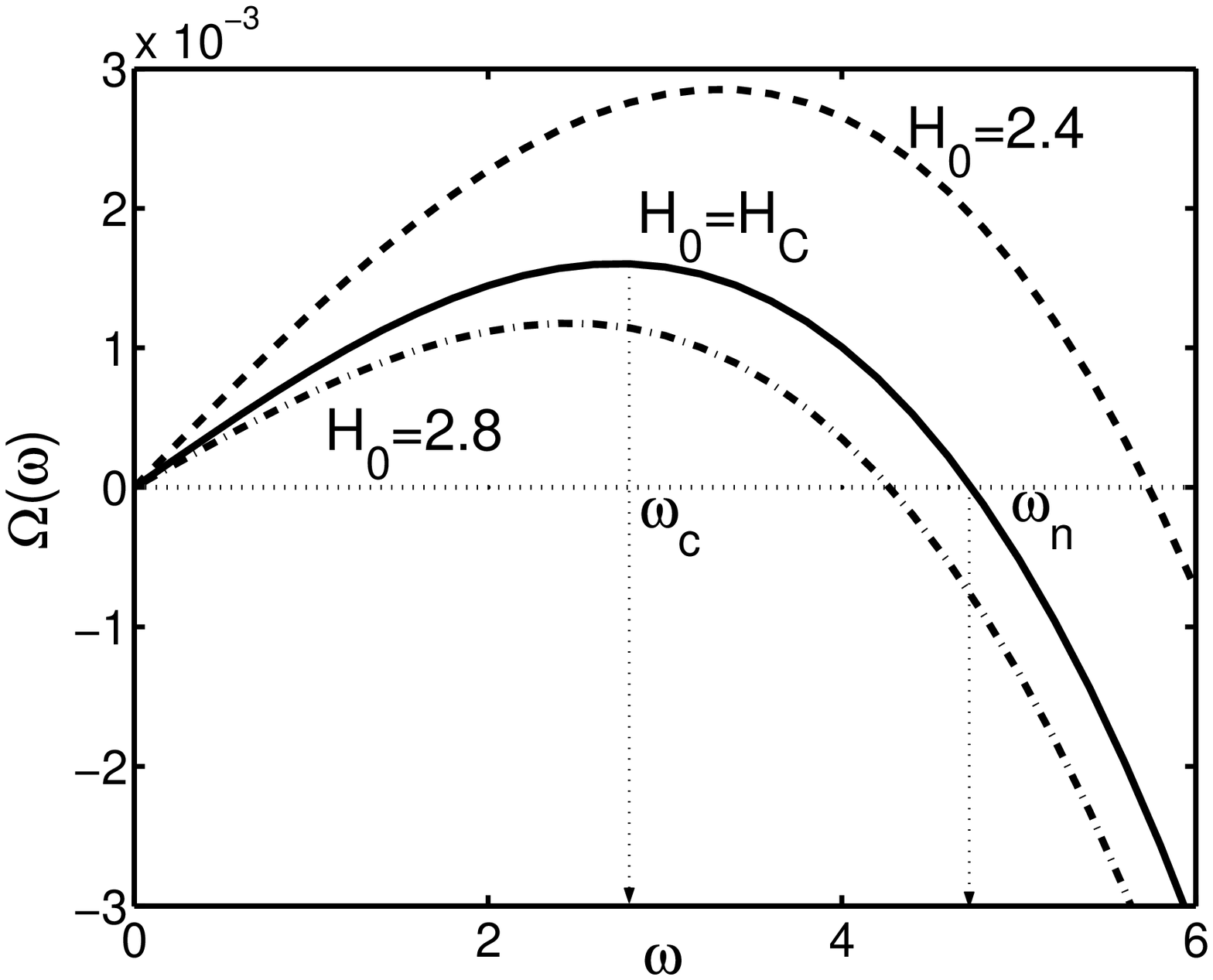}
\caption{\label{fig:epsart}
A plot of the growth factor $\Omega$ versus the wavenumber $\omega$ for three
values of the aspect ratio $c$ (plot on left), 
and for three values of the inclusion-interface distance $H_0$
(plot on right); $\omega_c$ is the critical
wavenumber, Eq. (31), and $\omega_n$ is the neutral mode, i.e. 
$\Omega(\omega_n)=0$. The critical distance, $H_c$, is represented by a
continuous line on the plot on the right.} 
\end{figure}
The maximum growth rate occurs at the critical wavenumber $\omega_c$ that is
obtained by maximizing $\Omega(\omega)$ with respect to $\omega$. We find,
\begin{equation}
\omega_c^2 = -{2 F(0) + {\cal G} \over 3(1+k) \sigma },
\end{equation}
with the understanding that the parameter values are such that the expression
in Eq. (31) is positive, i.e. $F(0) <0$ and $(2 F(0) + {\cal G})<0$.
On substituting Eq. (31) into the expression for the growth rate,
$\Omega(\omega)$, we obtain  the stability condition, $ {\cal G} < - 2 F(0)$,
and on using Equation (27), we find the following stability criterion for an
elliptic cylinder, valid for $c \approx 1$,
\begin{equation}
H_C \approx \sqrt{ { \pi^2 c(1+c) G_L (1-\alpha) \over
           6 (c+\alpha) {\cal G}    }
             },
\end{equation}
which reduces to the following exact criterion for a circular cylinder,
\begin{equation}
H_C = \sqrt{ { \pi^2 G_L (1-\alpha) \over
           3 (1+\alpha) {\cal G}    } }.
\end{equation}
The critical distance
$H_C$ is the largest value of $H_0$ for the instability to appear.
Note that the stability conditions, Eqs. (32)-(33), require that (i)
the product $G_L(1-\alpha)>0$, and (ii) $H_C >1$. Therefore, the instability
is observable only for certain combinations of the physical and processing
parameters.
\begin{figure}
\includegraphics[height=2.4in]{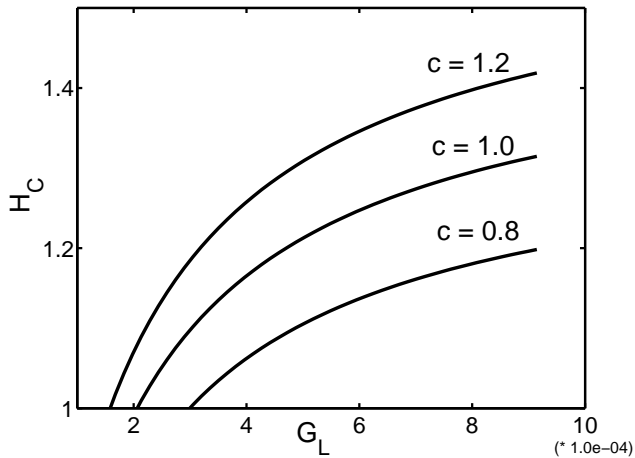}
\caption{\label{fig:epsart}
Plot of the critical inclusion-interface distance, $H_C$ (Eq. (32)), as 
function of the thermal gradient $G_L$ (note the scale factor $1.0e-04$)
 for three distinct values of the aspect ratio 
$c$, $c=1.2$, $c=1.0$ and $c=0.8$, for a system consisting of succinonitrile
(SCN) containing an inclusion with $(\alpha=5)$. The regions of instability 
in the $H_0-G_L$ plane are below the curves.
The numerical values of the parameters used in the calculations
are taken from Ref. \cite{Ste95}.}
\end{figure}
Figure 4 depicts the plot of $H_C$ versus the dimensionless thermal gradient
in the liquid for three distinct values of $c$. The numerical values of the
parameters used in calculating $H_C$ pertain to an experimental system 
consisting of SCN containing an inclusion whose thermal conductance ratio,
$\alpha$, is arbitrarily set equal to $0.5$.  We note the following:
(i) $H_C$ increases with the aspect ratio $c$, (ii) $H_C$ increases with
$G_L$, and (iii) for every value of the aspect ratio $c$, there exists a lower
bound for $G_L$ below which the instability is not observable, i.e.
$H_C<1$.  The dependence of $H_C$ on the inclusions size, i.e.
$a$, and on other factors, such as $\alpha$, resemble that of the case of
a spherical particle, with the exception that for spherical inclusions,
$H_C \sim (G_L/{\cal G})^{1/3}$. We refer the interested reader to 
\cite{Had04} for more details.

\end{document}